\documentclass[twocolumn,showpacs,superscriptaddress,english]{revtex4}
\usepackage{graphicx}

\makeatletter

\begin{document}

\title{Vulnerability and Hierarchy of Complex Networks}

\author{V. Gol'dshtein}
\email{vladimir@bgu.ac.il} \affiliation{Department of Mathematics,
Ben Gurion University of the Negev, Beer Sheva 84105, Israel}

\author{G.A. Koganov}
\email{quant@bgu.ac.il} \affiliation{Physics Department, Ben
Gurion University of the Negev, P.O.Box 653, Beer Sheva 84105,
Israel}

\author{G.I. Surdutovich}
\email{gregory@laser.nsc.ru} \affiliation{Laser Physics Institute,
Siberian Branch of Russian Academy of Sciences, Novosibirsk,
Russia}

\date{September 11, 2004}

\begin{abstract}
We suggest an approach to study hierarchy, especially hidden one,
of complex networks based on the analysis of their vulnerability.
Two quantities are proposed as a measure of network hierarchy. The
first one is the system vulnerability $V$. We show that being
quite suitable for regular networks this characteristic does not
allow one to estimate the hierarchy of large random networks. The
second quantity is a relative variance $h$ of the system
vulnerability that allows us to characterize a "natural" hierarchy
level of random networks. We find that hierarchical properties of
random networks depend crucially on a ratio $\delta$ between the
number of nodes and the number of edges. We note that any graph
with a transitive isometry group action (i.e. an absolutely
symmetric graph) is not hierarchical. Breaking such a symmetry
leads to appearance of hierarchy.
\end{abstract}

\pacs{(89.75.-k); (89.75.Fb); (89.75.Hh)}

\maketitle

\emph{$\quad$Introduction.} It is a traditional feeling that any
practical complex system (network) bears, at some degree, a
hierarchical property, which means that different parts (elements,
clusters, etc.) of the system have different impact on the system
performance. The very existence of hierarchy in the system can be
both explicit and implicit which do not necessarily coincide;
implicit hierarchy can even prevail over the explicit one, as is
the case e.g. with \char`\"{}gray cardinal\char`\"{} who can be
really much more powerful than the king. Intuitively, it seems
that the higher degree (the number of connections to the others) a
vertex has, the higher position it occupies in the system
hierarchy. Such type of hierarchy we call explicit hierarchy.
However, there are situations when vertices with maximal number of
edges are not necessarily most vital for the system performance.
For instance, all vertices in a binary tree have equal degree,
while there is strong hierarchy, i.e. the importance of a
particular vertex is dictated by the level it seats on: the
vertices which are closer to a root are more important than those
lying far from the root. Such type of hierarchy we will refer to
as a hidden hierarchy. In this letter we suggest a quantitative
way to recover a system hierarchy, especially implicit one, using
the system vulnerability properties.

The idea to relate the hierarchy and the vulnerability of the
system was inspired by a very simple reason that the more damage
can be caused by removal of a particular vertex, the higher
position in hidden hierarchy of the system this vertex occupies,
and vice versa. As a rough hierarchy measure we suggest a
pointwise version of the system vulnerability introduced by Latora
and Marchiori \cite{Latora-2004}, which is quite suitable for the
hierarchy characterization of regular systems. We will compare and
classify various regular networks with respect to their hierarchy.
As it will be demonstrated, the resulting hierarchical properties
can be quite different from an intuitive hierarchy. For randomized
networks the above described approach turned out to be
ineffective, therefore in order to quantify the hierarchy of
random networks we use statistical properties of the vulnerability
which are quite sensitive to the degree of hierarchy. It will be
shown that there exists a \char`\"{}natural\char`\"{} level of
hierarchy in randomized networks depending upon the ratio between
the number of vertices and the number of edges in the system.

We define a pointwise vulnerability $V(i)$ of the network as
relative drop in performance after removal of i-th vertex together
with all edges connecting it with other vertices, namely
\begin{equation}
V(i)=\frac{E-E(i)}{E}.\label{V(i)}\end{equation} \noindent Here
$E=\frac{1}{N(N-1)}{\textstyle \sum}_{i\neq j}\frac{1}{d_{ij}}$ is
the global efficiency \cite{Latora-2001} of the network, N is the
total number of vertices in the network, $d_{ij}$ is the minimal
distance (either weighted or unweighted) between the $i-th$ and
$j-th$ vertices, and $E(i)$ is the network efficiency after
removal of $i-th$ vertex and all its edges. Maximal value $V$ of
$V(i)$ corresponds to the network vulnerability introduced by
Latora and Marchiori \cite{Latora-2004}. As it was mentioned
above, we suggest to classify vertices of the network by the level
of their vulnerability $V(i)$. This seems to be a natural way to
introduce an hierarchy in any network by relating it to an ordered
distribution of vertices with respect to their vulnerability
$V(i)$. The most vulnerable vertex occupies the highest position
in the system hierarchy.

\begin{figure}
\centerline{\scalebox{0.45}{\includegraphics{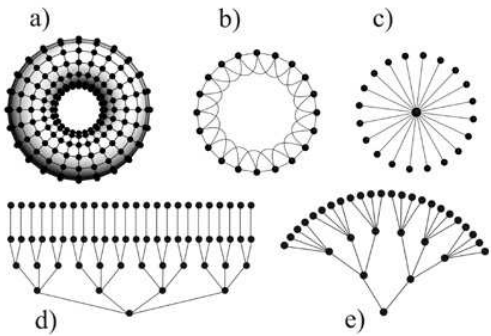}}}% Here is how to import EPS art
\caption{\label{Graphs} Regular networks: a) torus-like, b)
ring-like, c) spider-like, d) bush-like, e) tree-like.}
\end{figure}

\begin{figure}
\centerline{\scalebox{0.5}{\includegraphics{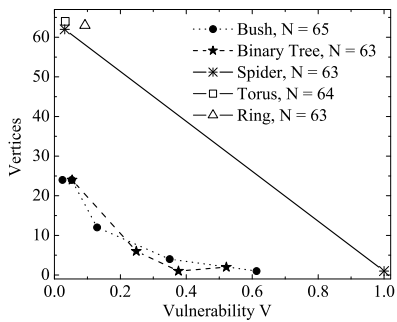}}}% Here is how to import EPS art
\caption{\label{reg-distrib} Vulnerability distributions for some
regular networks.}
\end{figure}

To illustrate our approach we have calculated the vulnerability of
several typical topologically different kinds of networks: a
tree-like, torus-like, ring-like, {}``bush''-like, and
{}``spider''-like networks, which are schematically shown in
Fig.\ref{Graphs}. The resulting vulnerability distributions $\rho$
are presented in Fig.\ref{reg-distrib}.

Two types of the considered systems, the torus-like and the
ring-like, are not hierarchical, which is caused by existence of
transitive action of an isometry group on these networks. In other
words, any vertex can be moved to any position by a suitable
one-to-one isometrical transformation. As a consequence, all
vertices in these networks have the same level of vulnerability,
i.e. $V(i)=const$. We will call any network with a transitive
action of isometry group an absolutely symmetric network. The
vulnerability distributions in such networks degenerate into a
single point as can be seen in Fig.\ref{reg-distrib} for both the
ring and the torus-like cases. Obviously, these systems are not
hierarchical as all their vertices have equal impact on overall
performance.

\emph{From this point of view absolute symmetry and hierarchy are
mutually excluding properties.}

In contrast to absolutely symmetric networks, the spider-like
networks have as strong as possible hierarchy, as the removal of a
central vertex gives rise to a complete destruction of the
network. Quantitatively it means that the vulnerability $V$ of
such system equals to one. The vulnerability distribution in this
case consists of two points (see Fig.\ref{reg-distrib}), one of
which corresponds to a central vertex with $V=1$, while the second
point is related to all other vertices with small value of
$V_{i}$. Notice that the symmetry of a spider-like system is
broken only in a single point, the central one. This prompts us to
conclude that the level of symmetry is not directly related to the
level of hierarchy. Indeed, the spider-like network is both quite
symmetric and highly hierarchical, but it is not absolutely
symmetric. It seems that interrelation between symmetry and
hierarchy is not so simple. The last two examples shown in
Fig.\ref{reg-distrib}, the tree-like and the bush-like ones,
demonstrate another possible type of hierarchy.

Our concept of hierarchy essentially differs from that proposed by
Trusina et al \cite{Trusina}. According to their definition the
bush-like network (Fig.\ref{Graphs}d) is maximally hierarchical,
while the tree-like one (Fig.\ref{Graphs}e) is maximally
antihierarchical. However as one can see on Fig.\ref{reg-distrib},
both these kinds of networks have quite similar hierarchical
properties. Moreover, intuitively the tree-like networks represent
typically hierarchical systems.

\begin{figure}
\centerline{\scalebox{0.5}{\includegraphics{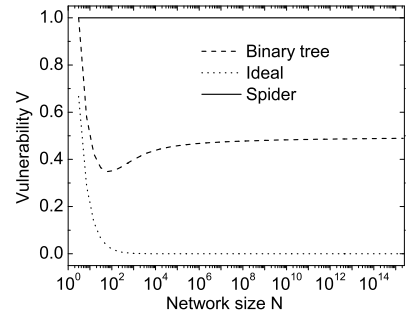}}}% Here is how to import EPS art
\caption{\label{V(N)} Vulnerability of regular networks as a
function of system size.}
\end{figure}

It is interesting to follow the dependence of vulnerability upon
the network size (graph order). Such dependencies for regular
networks are shown in Fig.\ref{V(N)} for the spider, an ideal
network (in which all pairs of vertices are connected) and a
binary tree. The first two exhibit opposite behavior. The
spider-like graph has maximal vulnerability $V=1$ at any size. On
the other hand, the vulnerability of the ideal network tends to
zero as the number of vertices grows. The vulnerability of any
other network is positioned between the vulnerability values of
the spider-like and of the ideal networks of the same size,
because any graph is a subgraph of the complete one. This fact is
illustrated by the example of binary tree (see Fig. \ref{V(N)}).

\begin{figure}
\centerline{\scalebox{0.4}{\includegraphics{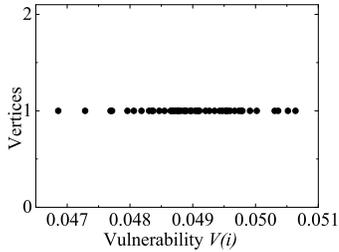}}}% Here is how to import EPS art
\caption{\label{rand-distrib} Vulnerability distribution for
randomized torus-like network.}
\end{figure}

So far we have considered non-random models of networks. Now we
will turn to randomized versions of networks and look what happens
with the vulnerability $V$ under randomization of the system. To
randomize a network we use the standard procedure of rewiring
randomly chosen pairs of vertices \cite{WS1998} (note that (i)
even a subtle difference in randomization procedure may result in
essentially different system behavior, as discussed in
\cite{our-0402338}; (ii) our procedure is not restricted by quite
strong condition of preserving the degree of every individual
vertex \cite{Trusina}). At the next step the vulnerability
distribution $\rho$ is calculated for every single realization,
afterwards a distribution $<\rho>$ of the random network is
obtained by averaging $\rho$ over all statistical realizations.
Now the vulnerability $V$ of the randomized network can be
introduced in the same way as it has been done earlier for regular
systems. The only difference is that now $V$ is defined from the
distribution $<\rho>$ rather than from $\rho$. As an example the
distribution $<\rho>$ for randomized spider-like graph is
presented in Fig.\ref{rand-distrib}.
This distribution has two essential differences from those for
regular networks (compare with Fig.\ref{reg-distrib}): (i) all
vulnerability values are within quite narrow range, (ii) all
vertices have different vulnerability values $V(i)$ in contrast to
the regular graphs where very large groups of vertices may have
the same value of $V(i)$.
\begin{figure}
\centerline{\scalebox{0.5}{\includegraphics{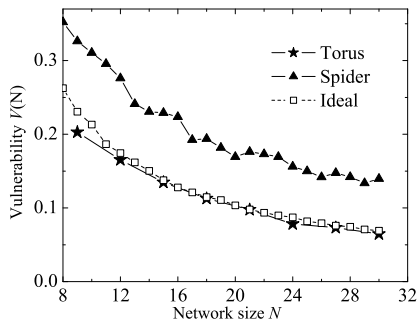}}}% Here is how to import EPS art
\caption{\label{rand-V(N)} Vulnerability of randomized networks as
a function of system size.}
\end{figure}
The vulnerabilities of totally randomized counterparts of the
spider-like, the torus-like and the ideal networks are plotted in
Fig.\ref{rand-V(N)} as functions of the system size. Surprisingly,
all three curves tend to zero with system size growing to
infinity, without any distinction between so different kinds of
network. Thus the asymptotic behavior of regular structures is
entirely different than that of their totally randomized
counterparts. To follow this drastic change we have calculated the
vulnerability $V$ of the same networks gradually increasing the
degree of randomization $p$, i.e. the fraction of randomly rewired
pairs of vertices from zero to one.
\begin{figure}
\centerline{\scalebox{0.5}{\includegraphics{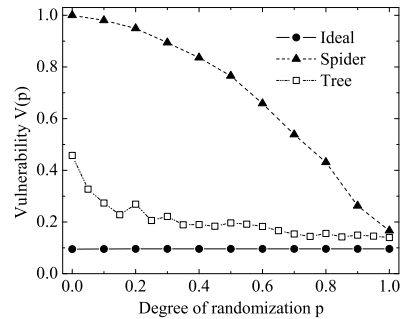}}}% Here is how to import EPS art
\caption{\label{V(p)} Vulnerability of randomized networks as a
function of randomization degree p.}
\end{figure}
As one can see from Fig.\ref{V(p)}, the impact of randomization on
the vulnerability depends crucially on the initial network
structure. Being essentially different at no randomization limit
$p=0$ which corresponds to the regular case (see Fig.\ref{V(N)}),
all three types of networks become indistinguishable at the
totally random limit $p=1$. We conclude that the vulnerability
$V(i)$ is not a suitable structural characteristic of random
networks because from this point of view large random networks are
not hierarchical at all, which seems to be doubtful.

As it was already mentioned, the value $<V>$ of the mean
vulnerability may also be used to characterize the network
hierarchy level. However this quantity also tends to zero with
decrease of the system size, as well as the vulnerability $V$.
Generally, both the mean and the maximal parameter values are very
rough characteristics of large random systems. For example, the
mean ocean level is constant. From this point of view the ocean is
rather homogeneous system without any hierarchy. However this is
not correct as there are always highly hierarchical storm regions.
Another example is the mean temperature of patients in a hospital
that tells nothing about the local situations. To obtain more
detailed description one has to use other parameters which are
more sensitive to deviations from the mean value, as it is usual
for statistics. As such additional parameter we introduce a
relative variance $h$ of the pointwise vulnerability $V_{i}$ as a
measure of a random network hierarchy:
\begin{equation}
h=\frac{<\Delta V^{2}>}{<V>^{2}},
\end{equation} \noindent where
$V$ stands for the vulnerability of a single statistical
realization of the network, $<\Delta
V^{2}>=\frac{1}{N}\sum_{i=1}^{N}(V(i)-<V>)^{2}$, and
$<V>=\frac{1}{N}\sum_{i=1}^{N}V_{i}$ is the mean pointwise
vulnerability. The parameter $h$ is a measure of the fluctuation
level and, as will be seen, can be used to describe the
hierarchical properties of both regular and random networks. This
means that in the case of ocean mentioned above wave intensities
should be analyzed instead of wave amplitudes in order to find the
hierarchical structure.

\begin{figure}
\centerline{\scalebox{0.5}{\includegraphics{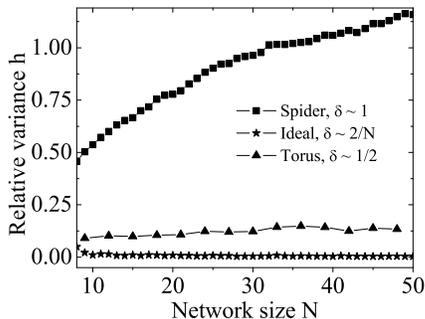}}}% Here is how to import EPS art
\caption{\label{h(N)} Degree of hierarchy $h$ of random networks
as a function of the system size.}
\end{figure}

The dependence of $h$ upon the system size for the randomized
versions of the spider-like, the torus-like, and the ideal-like
types of network is presented in Fig. \ref{h(N)}. In contrast to
asymptotic behavior of the vulnerability $V$ that tends to zero at
large size limit for all three networks, the asymptotic behavior
of the relative variance $h$ exhibits a crucial difference for
different types of networks. For the ideal network $h\rightarrow0$
which means that the level of fluctuations remains extremely low
despite the total randomization of the network. The situation
changes drastically for the spider-like system where the relative
variance $h$ grows with the system size and becomes even bigger
than one for large enough systems, i.e. the fluctuations are huge
in spider-like systems. Again we have two extreme situations,
namely $h\rightarrow0$ for the ideal and $h\sim1$ or bigger for
spider-like random networks (compare with Fig.\ref{V(N)}). The
value of $h$ for the randomized torus is between these two extreme
values.

What makes totally random networks of the same size $N$ so
different? What is their {}``memory'' about their non-randomized
parents? We believe that the role of such a relict factor is
played by the relative number of connections (edges), therefore to
classify different types of random networks according to
asymptotic behavior of their fluctuations $h$ we introduce the
ratio $\delta$ of the number of vertices $N$ to the number of
edges. This quantity equals to $N/(N-1)$ for the tree and the
spider-like graphs and $\delta=2/(N-1)$ for the ideal network. It
should be noted that the randomization procedure used here
strictly preserves $\delta$. As it can be seen in Fig. \ref{h(N)},
the level of fluctuations characterized by the value of $h$
increases with $\delta$, which is clear since the larger $\delta$
the lesser the number of edges and so there are more options for
randomization. Thus, the relative variance $h(\delta)$ introduced
above can be used as the hierarchy measure for random networks.

\emph{Summary and discussion}. We have introduced the
vulnerability distributions for complex networks and used them for
studying the network hierarchical structure. Such distributions
are essentially different for different types of networks and can
be quite complicated for rigorous analysis. To estimate the
hierarchy level of networks, both regular and random, we have used
the following global characteristics: the network vulnerability
$V$ and the relative variance $h$ of the poitnwise vulnerability
$V(i)$. It has been demonstrated that the relative variance is a
more delicate hierarchy characteristic than the vulnerability,
especially for random networks in which the hierarchy level $h$
depends crucially upon the ratio $\delta$ of the number of
vertices to the number of edges. Any random network has its
natural hierarchy level $h(\delta)$ which is minimal for the ideal
random network with $\delta\rightarrow0$ and is maximal for the
spider-like random network with $\delta\sim1$ (strictly speaking,
one can imagine networks with the number of edges lesser than that
in the spider-like graph, i.e. $\delta>1$, but such networks are
disconnected, and this situation is beyond the scope of this
letter). This prompts us to make a more general conclusion that
stochasticity itself does not exclude the presence of hierarchy,
and any randomized system can have some local islands with quite
hierarchical structures. Complexity of such hierarchy islands is
higher for systems with restricted number of connections.

Our approach is fully applicable when a link hierarchy is of
interest rather than hierarchy of vertices. Then the quantity
$E(i)$ in eq.(\ref{V(i)}) would mean the network efficiency after
removal of $i-th$ link, hence eq.(\ref{V(i)}) will define a
linkwise vulnerability \cite{Surd}. Hierarchical structures based
on pointwise and linkwise vulnerabilities of the same network can
look entirely different. The obvious example is the spider-like
network where the pointwise based approach results in strongest
possible hierarchy ($V=1$), whereas the linkwise
 vulnerability tends to zero in large size limit, meaning that the
 there is no hierarchy of links whatsoever.

 Although all systems considered here are idealized mathematical
models, both regular and random ones, our approach is applicable
for analysis of the vulnerability $V$ and the degree of hierarchy
$h$ of any real network. Finally, apart from $V$ some other
hierarchy measures based upon vulnerability distributions may be
used, such as mean vulnerability, the number of vulnerability
levels, etc.

\end{document}